\documentclass[useAMS,usenatbib]{mn2e}
\usepackage{natbib}
\usepackage{graphicx}
\usepackage{times}
\usepackage{rotate}
\usepackage{color,url}
\newcommand{\prd}{Phys.~Rev.~D}

\newcommand{\mnras}{MNRAS}
\newcommand{\apj}{ApJ}
\newcommand{\aj}{AJ}

\newcommand{\physrep}{Physics Report}
\newcommand{\nat}{Nature}
\newcommand{\jcap}{JCAP}

\newcommand{\simgt}{\lower.5ex\hbox{$\; \buildrel > \over \sim \;$}}
\newcommand{\simlt}{\lower.5ex\hbox{$\; \buildrel < \over \sim \;$}}

\begin{document}
\title[HOD and Growth Rate from Multipole Power Spectrum]
{
Constraining Halo Occupation Distribution and Cosmic Growth Rate using Multipole Power Spectrum
}

\author[Hikage]{Chiaki Hikage$^1$ \\
$^1$ Kobayashi-Maskawa Institute for the Origin of Particles and the Universe (KMI), Nagoya University, 464-8602, Japan}
\maketitle

\label{firstpage}

\begin{abstract}
We propose a new method of measuring halo occupation distribution
(HOD) together with cosmic growth rate using multipole components of
galaxy power spectrum $P_l(k)$.  The nonlinear redshift-space
distortion due to the random motion of satellite galaxies, i.e.,
Fingers-of-God, generates high-$l$ multipole anisotropy in galaxy
clustering such as the hexadecapole ($l=4$) and tetra-hexadecapole
($l=6$), which are sensitive to the fraction and the velocity
dispersion of satellite galaxies. Using simulated samples following
the HOD of Luminous Red Galaxies (LRGs), we find that the input HOD
parameters are successfully reproduced from $P_l(k)$ and that high-$l$
multipole information help to break the degeneracy among HOD
parameters.  We also show that the measurements of the cosmic growth
rate as well as the satellite fraction and velocity dispersions are
significantly improved by adding the small-scale information of
high-$l$ multipoles.
\end{abstract}

\begin{keywords}
cosmology: theory -- observations -- large-scale
structure of the Universe -- galaxies: kinematics and dynamics
\end{keywords}

\vspace{-0.5cm}

\section{Introduction}
\label{sec:intro}
Understanding the relationship between galaxy distributions and their
host dark matter halos is a key ingredient in the physics of galaxy
formation and is also important for precision cosmology study using
galaxy datasets. Halo occupation distribution (HOD) describes the
probability distribution of the occupation number of central and
satellite galaxies as a function of their host halo mass
\citep[e.g.,][]{BerlindWeinberg02,Berlind03,Kravtsov04,Zheng05}. Abundance matching of
simulated subhalos to connect the properties of galaxies has been
widely investigated \citep[e.g.,][]{Masaki13}. HOD has been measured
from a variety of galaxy samples using projected correlation function
which is sensitive to the radial profile of galaxies
\citep[e.g.,][]{Zehavi05,Masjedi06,Zheng09,White11,Geach12} and also
from galaxy group multiplicity functions[e.g.,][]{Reid09a}. Cross
correlation of galaxies with background galaxy image distortions,
i.e., galaxy-galaxy lensing, has been measured to connect the halo
mass with the galaxy properties \citep[e.g.,][]{Mandelbaum06}. The
dilution of the lensing signal around off-centered (satellite)
galaxies provides a probe of the satellite fraction and the velocity
dispersion \citep{Hikage12,George12,Hikage13}.

We propose a novel method to constrain HOD using the multipole galaxy
power spectra $P_l(k)$ by characterizing the anisotropy of the galaxy
clustering due to the redshift-space distortion (RSD).  The random
motion of galaxies inside their host halos generates the nonlinear
redshift-space distortion, i.e., Fingers-of-God (FoG) effect
\citep{Jackson72}. As the off-centered or satellite galaxies have
large internal motion, the FoG effect provides a useful probe of
constraining the fraction of satellite galaxies.  \cite{HY2013} report
a clear detection of high-$l$ multipole anisotropy such as
hexadecapole ($l=4$) and tetra-hexadecapole ($l=6$), which comes from
the FoG effect of one-halo term, the contributions of
central-satellite and satellite-satellite pair hosted by the same
halos.

In this letter, we utilize such high-$l$ multipole anisotropy to
constrain HOD. Focused on the luminous red galaxies (LRGs), which has
been widely used for cosmology analysis in SDSS \citep{Eisenstein01},
we construct simulated mock LRG samples on the HOD basis.  We show
that the input HOD is reproduced from $P_l(k)$.  High-$l$ multipole is
sensitive to satellite velocity distribution, while the projected
correlation function depends on the radial profile of satellite
galaxies in the host halos. Both measurements play complimentary roles
in understanding the relationship between galaxies and halos.

The bulk motion of galaxies drives the anisotropy in galaxy
clustering, which provides a powerful observational probe of measuring
the cosmic growth rate to test General Relativity and various gravity
models \citep{Peacock01,Okumura08,Guzzo08}. Combinations of monopole
($l=0$) and quadrupole ($l=2$) power spectra has been widely used to
constrain growth rate from various galaxy samples including SDSS LRG
\citep{Yamamoto08,Yamamoto10,Sato11,Oka13} and recently BOSS CMASS
sample \citep{Reid12,Beutler13,Samushia13}.  Satellite FoG effect is a
major systematic uncertainty in measuring the growth rate in this
analysis. Even when the satellite fraction of the SDSS LRG sample is
just 6\%, the measurement of growth rate can be strongly biased
\citep{HY2013}.  As high-$l$ multipole spectra such as $P_4$ and $P_6$
are sensitive to the satellite fraction, they are useful for
eliminating the uncertainty of the satellite FoG. In this letter, we
show that the measurement of growth rate as well as HOD parameters is
significantly improved by adding high-$l$ multipole information.

This letter is organized as follows: in section 2, we present the
theoretical formalism to describe the multipole power spectra based on
the halo model.  In section 3, we summarize how to make simulation
data for LRG catalogs. The results of the measurements of HOD and the
growth rate are summarized in section 4.  Section 5 is devoted to
summary and conclusions. Throughout the letter, we use flat $\Lambda$
CDM model with the fiducial cosmological parameters as follows:
$\Omega_bh^2=0.0226$, $\Omega_ch^2=0.1108$, $n_s=0.963$, $h=0.704$,
$\tau=0.089$, $\sigma_8=0.817$.

\vspace{-0.5cm}

\section{Formalism}
\label{sec:formalism}
In this section, we summarize our theoretical formulae of the
multipole power spectra of LRGs based in the HOD framework.

\vspace{-0.5cm}
\subsection{Multipole power spectra in halo model}
Galaxy power spectrum in redshift space $P(k,\mu)$, where $\mu$ is the
cosine of the angle between the wavevector $\mathbf{k}$ and the line-of-sight direction, is
described by expanding their multipole components:
\begin{equation}
\label{eq:pkl}
P_l(k)=\frac{2l+1}{2}\int_{-1}^1 d\mu P(k,\mu) {\cal L}_l(\mu),
\end{equation}
where ${\cal L}_l$ is $l$-th Legendre polynomials. In the halo-model
approach \citep[e.g.,][]{Seljak01,White01,CooraySheth02}, the galaxy power
spectrum can be decomposed into the one-halo and two-halo terms,
\begin{equation}
P(k,\mu)=P^{\rm 1h}(k,\mu)+P^{\rm 2h}(k,\mu).
\label{eq:pklmodel}
\end{equation}
One-halo term is the contribution from the clustering of
central-satellite and satellite-satellite pairs hosted by same halos
and written as follows:
\begin{eqnarray}
P^{\rm 1h}(k,\mu)&=&\frac{\displaystyle 1}{\displaystyle n_{\rm tot}^2}
\int dM\frac{dn_{\rm h}}{dM}
\Bigl[2\langle N_{\rm cen} N_{\rm sat}\rangle
\tilde{p}_{\rm sat}(k,\mu; M) \nonumber \\
&&+\langle N_{\rm sat}(N_{\rm sat}-1)\rangle \tilde{p}_{\rm sat}^2(k,\mu; M)\Bigr], 
\label{eq:pk_1h}
\end{eqnarray}
where $dn_{\rm h}/dM$ is the halo mass function, $\langle N_{\rm
  cen}\rangle$ and $\langle N_{\rm sat}\rangle$ denote the HOD of
central and satellite galaxies respectively, $n_{\rm tot}$ is the
total number density of galaxies (see the details of HOD
parametrization in the next subsection).  Here we assume that the halo
hosting satellite LRGs must have a central LRG and thereby $\langle
N_{\rm cen}N_{\rm sat} \rangle=\langle N_{\rm sat}\rangle$
\citep{Zheng05}. This assumption is good for our purpose, though some
of central galaxies are not always LRGs \citep[e.g.,][]{Skibba11}. We
assume that the occupation number of satellite galaxies follows
Poisson statistics, which implies $\langle N_{\rm sat}(N_{\rm
  sat}-1)\rangle=\langle N_{\rm sat} \rangle^2$
\citep{Kravtsov04,Zehavi05}.  We neglect the velocity dispersion of
central galaxies. The satellite distribution around central galaxies
inside the halo with mass $M$ is described with $\tilde{p}_{\rm
  sat}(k,\mu;M)$.  We consider that the radial profile of satellite
galaxies follows the NFW profile \citep{NFW} with the concentration
given by \cite{Duffy08} and the internal velocity of satellite
galaxies has Gaussian distribution
\begin{eqnarray}
\tilde{p}_{\rm sat}(k,\mu,M)&=&\tilde{u}_{\rm NFW}(k;M)
\exp\left[-\frac{\sigma_v^{\rm (vir) 2}(M)k^2\mu^2}{2a^2H^2(z)}\right].
\label{eq:psoff2}
\end{eqnarray}
where $\tilde{u}_{\rm NFW}(k)$ is the Fourier transform of the NFW
density profile.  The velocity dispersion of satellite galaxies is
given by the Virial velocity dispersion $\sigma_v^{\rm (vir)}\equiv
(GM/2r_{\rm vir})^{1/2}$ and then the average velocity dispersion of
satellite galaxies is given using the satellite HOD as
\begin{equation}
\sigma_v^{\rm (sat)}=\left[\frac{1}{n_{\rm sat}}\int dM\frac{dn_{\rm h}}{dM}
\langle N_{\rm sat}\rangle \sigma_v^{\rm (vir) 2}\right]^{1/2},
\label{eq:vsat}
\end{equation}
where $n_{\rm sat}$ is the number density of satellite galaxies.  The
satellite fraction $f_{\rm sat}$ is defined as $n_{\rm sat}/n_{\rm
  tot}$. 

The two-halo term, which is the contribution of the clustering of LRGs
in different halos, depends on the redshift-space halo power spectra
with different halo masses $P_{\rm hh}(k;M,M')$ \citep{Hikage13}:
\begin{eqnarray}
& P^{\rm 2h}(k,\mu)=\frac{\displaystyle 1}{\displaystyle n_{\rm tot}^2}
\int dM\frac{\displaystyle dn_{\rm h}}{\displaystyle dM} 
\int dM'\frac{\displaystyle dn_{\rm h}}{\displaystyle dM'}~~~~~~~~~~~~~~~~~~~~~~~~~~~
\nonumber \\
& \times\left[\langle N_{\rm cen}\rangle+\langle N_{\rm sat}\rangle\tilde{p}_{\rm sat}(k,\mu; M)\right]~~~~~~~~~~~~~~~~~~~~~~~
 \nonumber \\
& \times\left[\langle N_{\rm cen}\rangle +\langle N_{\rm sat}\rangle\tilde{p}_{\rm sat}(k,\mu; M')\right] 
P_{\rm hh}(k,\mu;M,M').
\label{eq:pk_2h}
\end{eqnarray}
The two-halo term is also affected by FoG effect of satellite galaxies
while the internal motion of central galaxy is assumed to be
negligible. We directly estimate the redshift-space halo power spectra
$P_{\rm hh}(k,\mu;M,M')$ using simulations to incorporate
various nonlinear effects such as the gravitational evolution, the halo
biasing, and the halo motion. We divide the simulated halo samples
into 10 different mass bins to compute their auto- and cross-halo
power spectra in redshift space (see section \ref{sec:3} for details).

\vspace{-0.5cm}

\subsection{Parametrization of Halo Occupation Distribution (HOD)}
We consider two different ways to describe HOD of central and
satellite LRGs. 

\begin{enumerate}
\item One is using a following functional form of HOD
\citep{Zheng05}:
\begin{eqnarray}
\langle N_{\rm cen}\rangle &=&\frac{1}{2}\left[1+{\rm erf}\left(\frac{\log_{10}(M)-\log_{10}
(M_{\rm min})}{\sigma_{\log M}}\right)\right], \nonumber \\
\langle N_{\rm sat}\rangle &=&\langle N_{\rm cen}\rangle\left(\frac{M-M_{\rm cut}}{M_1}\right)^{\alpha},
\label{eq:HOD}
\end{eqnarray}
where ${\rm erf}(x)$ is the error function and five HOD parameters are
included. The fiducial values of the HOD parameters for SDSS DR7 LRG
sample is set as $M_{\rm min}=5.7\times 10^{13}h^{-1}M_\odot$,
$\sigma_{\log M}=0.7$, $M_{\rm cut}=3.5\times 10^{13}h^{-1}M_\odot$,
$M_1=3.5\times 10^{14}h^{-1}M_\odot$, and $\alpha=1$ \citep{Reid09a}.

\item 
The other way is not assuming any functional form of HOD. We
parametrize central and satellite HODs at different mass and
interpolate the HOD of intermediate halo mass with cubic spline
approximation.  We divide 5 different mass bins (10 HOD parameters in
total) in the mass range from $10^{12}h^{-1}M_\odot$ to
$10^{15}h^{-1}M_\odot$.  We assume that the satellite number $\langle
N_{\rm sat}\rangle$ monotonically increases as larger halo mass $M$.
\end{enumerate}

\vspace{-0.5cm}

\section{Simulation with the HOD of LRGs}
\label{sec:3}
We construct simulated samples with the HOD of LRGs to estimate the
error of HOD parameters and growth rate from the measurements of
$P_l$. We run 100 realizations of N-body simulations using Gadget-2
code \citep{Springel05}. The initial distribution of mass particles is
set using 2LPT code in Gaussian initial condition \citep{Crocce06}
with the initial redshift of $z=49$.  The initial matter power
spectrum is computed using CAMB software \citep{Lewis00} with the
fiducial cosmological parameters. The N-body simulations are performed
in a periodic cubic box at the side length $L_{\rm box}$ of
$1h^{-1}$Gpc with the number of mass particles is 800$^3$ where each
particle mass is $1.3 \times 10^{11}h^{-1}M_\odot$. Halo is identified
by Friends-of-Friends algorithm with the linking length $b=0.2$. The
minimum number of mass particles constituting halos is 20, which
corresponds to the halo mass of $2.6\times 10^{12}h^{-1}M_\odot$. The
position and velocity of each halo is defined as the arithmetic mean
of those of the constituent mass particles of each halo.

We randomly select halos hosting a central LRG and pick up the dark
matter particles for satellite LRGs to follow the fiducial HOD
(eqs. [7,8]). The number distribution of satellite LRGs follow
Poisson statistics. The position and velocity of the host halo and the
mass particles are respectively assigned to those of central and
satellite LRGs. In the fiducial HOD values, the fraction of satellite
LRGs is 6.5\% and their average velocity dispersion $\sigma_v^{\rm
  (sat)}$ is $570$km/s in average. For simplicity, the simulations do
not include various observational issues such as the angular mask,
radial selection function, and fiber collisions.

Number density is assigned to each grid with the nearest grid point
(NGP) scheme and the number of grids $N_{\rm grid}$ is set to be 512
at a side. We compute the multipole power spectrum by transforming the
density field into Fourier space with Fast Fourier Transform (FFT)
method.  The shot noise term, the inverse of the number density in the
sample, is subtracted from the monopole power spectrum. The covariance
of the multipole power spectra $P_l$ at different $l$ and different
bins of $k$ is estimated from 100 realizations of simulated samples.
We also combine jackknife resampling method and obtain the enough
number of realizations to estimate the covariance. 

We take z-axis as the line-of-sight direction and obtain the
redshift-space position by adding the velocity component $v_z$
\begin{equation}
s=r+\frac{f_z}{f_z^{\rm (GR)}}\frac{(1+z)v_z}{H(z)}.
\label{eq:zdist}
\end{equation}
In order to estimate the deviation of the growth rate $f_z$ from the
GR prediction $f_z^{\rm (GR)}$, we simply change the amplitude of
velocity by hand \citep[c.f.,][]{NishimichiOka13}.  We compute the
halo power spectrum for $f_z=(1\pm \Delta)f_z^{\rm (GR)}$ with
$\Delta=0.1$ and linearly interpolate them to estimate $P_{\rm hh}$
for arbitrary $f_z$ in the theoretical model (eq.[\ref{eq:pk_2h}]).
For simplicity, we neglect the change of the halo mass function and
the halo bias from their GR prediction.

\vspace{-0.5cm}

\section{Results}

\subsection{Satellite FoG effects on multipole power spectra}
\begin{figure}
\begin{center}
\includegraphics[width=4.1cm]{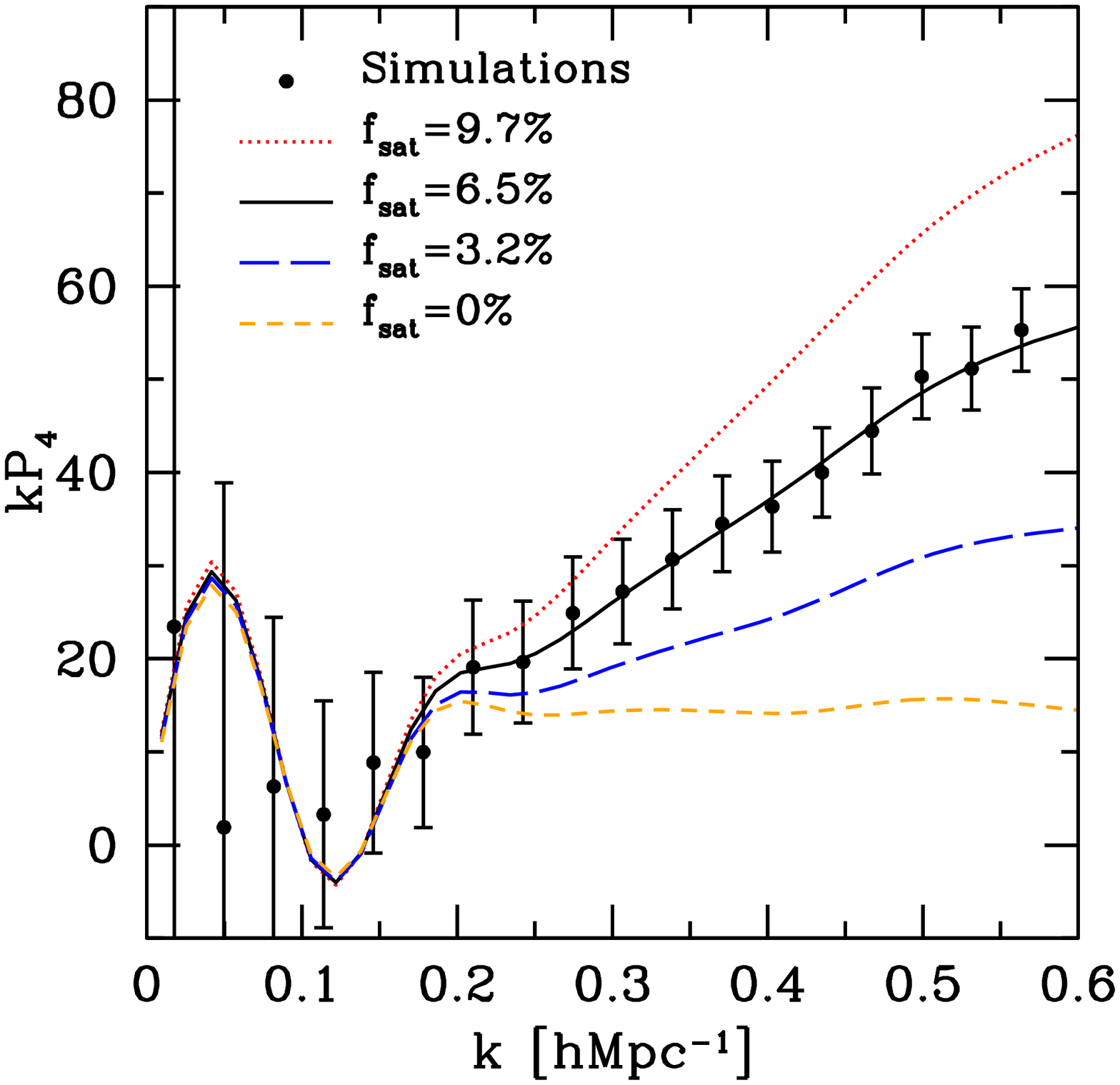}
\includegraphics[width=4.1cm]{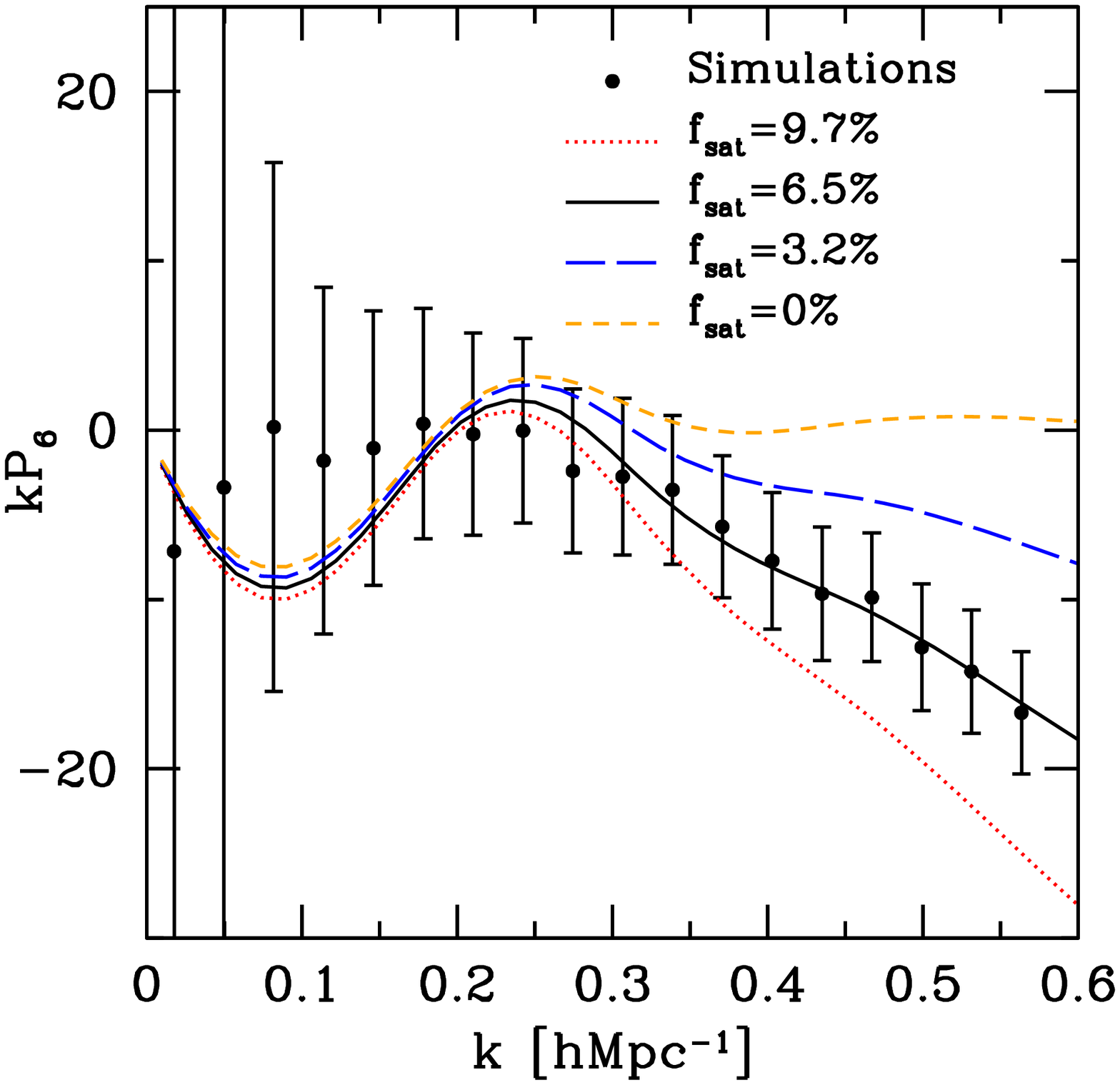}
\includegraphics[width=4.1cm]{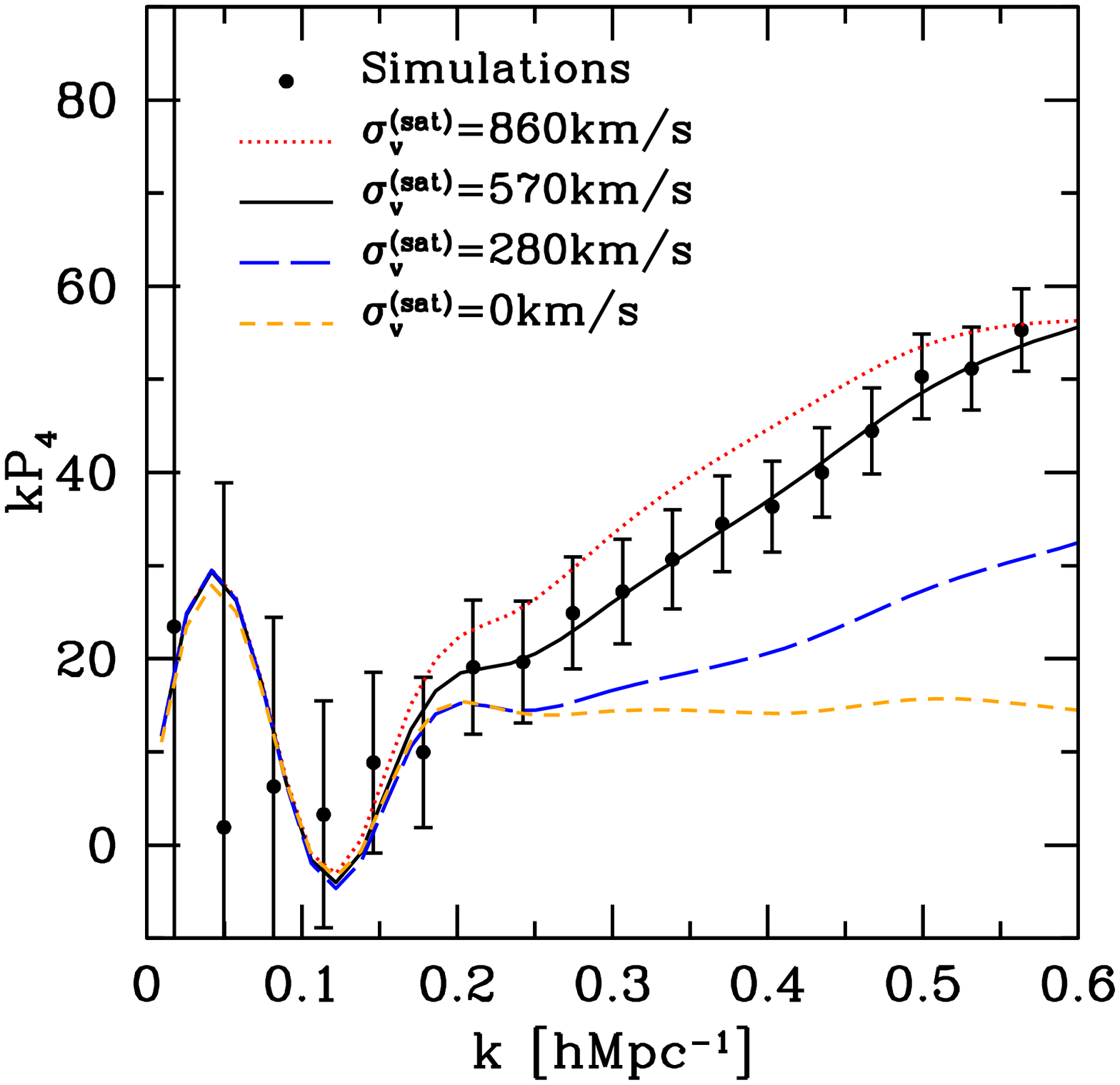}
\includegraphics[width=4.1cm]{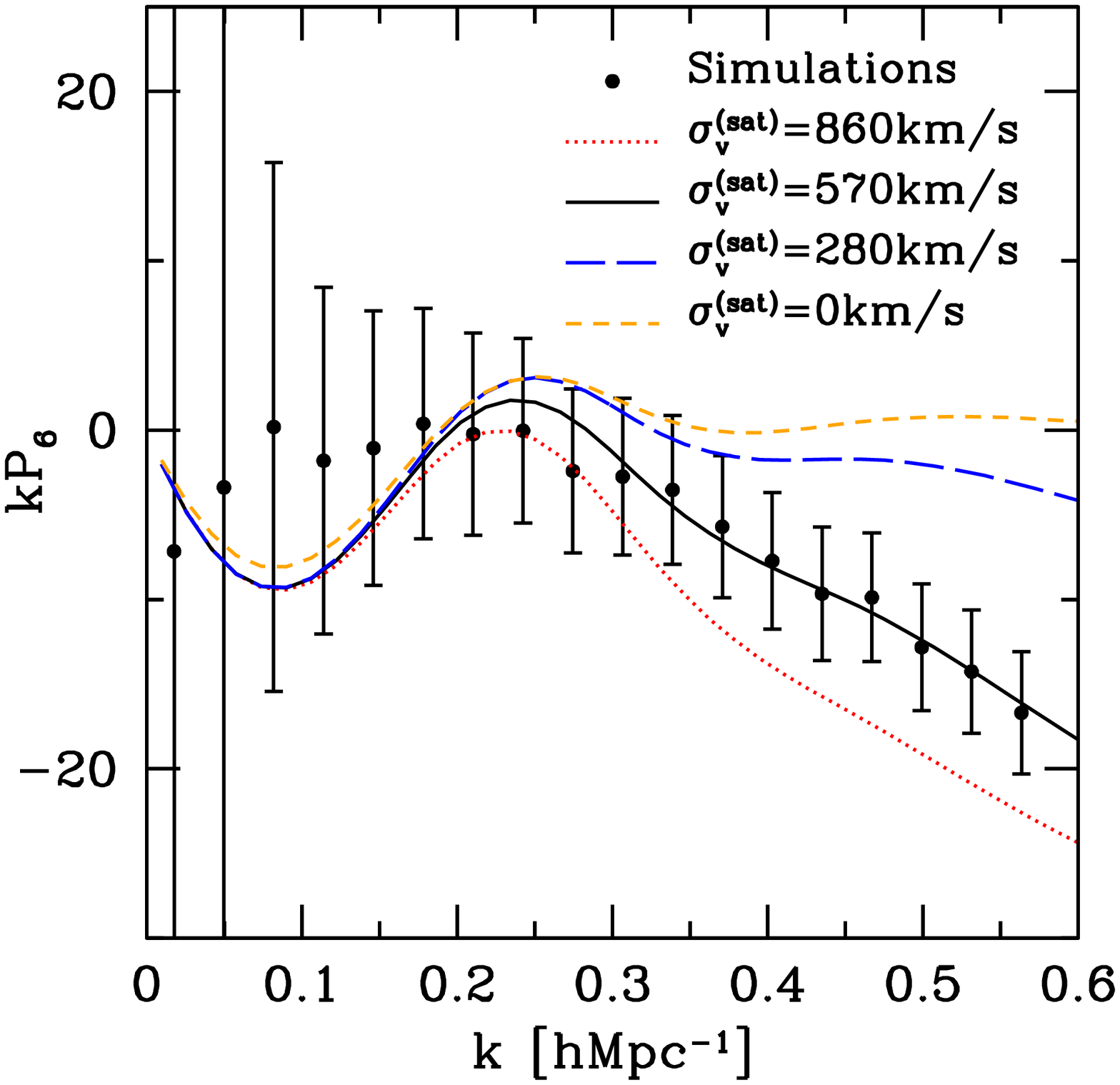}
\caption{Hexadecapole $P_4(k)$ (left) and tetra-hexadecapole $P_6(k)$
  (right) for the mock LRG samples in average (black
  circles). Error-bars denote the 1-$\sigma$ dispersion. The black
  lines represent the model prediction with the fiducial values of HOD
  in which the satellite fraction is $f_{\rm sat}=6.5$\% and the
  satellite velocity dispersion is $\sigma_v^{\rm (sat)}=570$km/s. For
  comparison, we plot the theoretical models by varying $f_{\rm sat}$
  (Upper) and $\sigma_v^{\rm (sat)}$ (Lower). The binning width
    $\Delta k$ for these plots is $0.032h$/Mpc, which is twice larger
    than that in the fitting.} \label{fig:pkl_fsat}
\end{center}
\end{figure}

Figure \ref{fig:pkl_fsat} shows the simulation results of $P_4$ and
$P_6$ (black circles) in comparison with the model predictions with
the same HOD (black lines). They are found to be in excellent
agreement with each other. For comparison, we plot the model
predictions by varying the satellite fraction $f_{\rm sat}$ and the
satellite velocity dispersion $\sigma_v^{\rm (sat)}$.  High-$l$
multipoles, which is mainly determined by the one-halo term, are
sensitive to the satellite properties through their velocity
distribution or FoG effect.  As the satellite fraction is small for
the LRG sample, central-satellite pair contribution is dominant and
thus the overall amplitude of $P_l$ ($l\ge 4$) is roughly proportional
to $f_{\rm sat}$ (see eq.[\ref{eq:pk_1h}]). As $\sigma_v^{\rm (sat)}$
increases, the FoG effect starts at smaller $k$ while the overall
amplitude of $kP_l^{1h}$ at large-$k$ limit decreases
\citep{HY2013}. The feature can be seen especially in $P_4$ in the
lower panels of Figure \ref{fig:pkl_fsat}.

\begin{table*}
\begin{center}
\begin{tabular}{cccccccccc}
\hline
\hline
 & & \multicolumn{5}{c}{\raisebox{0ex}{$\sigma_8$, $\delta N$ fixed}} & & \multicolumn{2}{c}{\raisebox{0ex}{$\sigma_8$, $\delta N$ free}} \\
\cline{3-7}\cline{9-10}
parameters & input & $P_{0,2} (k\le 0.2)$ & $P_{0,2} (k\le 0.2)$ & $P_{0,2} (k\le 0.2)$ & $P_{0,2} (k\le 0.6)$ & $P_{0,2} (k\le 0.6)$ & & $P_{0,2} (k\le 0.6)$ & $P_{0,2} (k\le 0.6)$ \\
           & values &  --              & $P_{4}   (k\le 0.6)$ & $P_{4,6} (k\le 0.6)$ & -- & $P_{4,6} (k\le 0.6)$ & & -- & $P_{4,6} (k\le 0.6)$ \\
\hline
$M_{\rm min}$     & 5.7 & $5.28\pm 0.36$ & $5.47\pm 0.23$ & $5.45\pm 0.22$ & $5.54\pm 0.11$ & $5.54\pm 0.11$ & & $6.37\pm 0.91$ & $5.52\pm 0.79$  \\
$\sigma_{\log M}$ & 0.7 & $0.66\pm 0.04$ & $0.68\pm 0.02$ & $0.68\pm 0.02$ & $0.68\pm 0.01$ & $0.68\pm 0.01$ & & $0.76\pm 0.14$ & $0.68\pm 0.07$  \\
$M_{\rm cut}$     & 3.5 & $ 6.2\pm 4.1$  & $ 4.8\pm 2.4$  & $ 4.4\pm 2.4$  & $ 1.4\pm  1.7$ & $ 3.4\pm  1.4$ & & $ 6.4\pm  3.4$ & $ 3.8\pm  2.0$  \\
$M_1$             & 35  & $  35\pm 13$   & $ 34\pm 6$     & $  34\pm 5$    & $  37\pm    3$ & $  35\pm  2.2$ & & $  35\pm    7$ & $ 34\pm   2.9$  \\
$\alpha$          & 1   & $0.96\pm 0.26$ & $0.90\pm 0.21$ & $0.94\pm 0.21$ & $1.04\pm 0.20$ & $1.02\pm 0.19$ & & $0.84\pm 0.24$ & $1.01\pm 0.22$  \\
\hline
$\chi^2_{\rm min}$ (d.o.f.)  & --  &  0.2 (20) & 1.7 (57) & 2.2 (94) & 1.0 (70) & 2.9 (144) & & 0.88 (68) & 2.8 (142) \\
\hline
\end{tabular}
\caption{The best-fit values of HOD parameters in the functional form
  (eq.[\ref{eq:HOD}]) measured from different combinations of $P_l$
  for $k\le 0.2$ or $0.6h$/Mpc ($l=0,2$) and $k\le 0.6h$/Mpc ($l=4,6$)
  averaged over 100 simulated samples.  Cosmological parameters are
  fixed except for the rightmost two columns, which shows the results
  including $\sigma_8$ and the residual shot noise $\delta N$ as
  additional free parameters. The input values of HOD parameters are
  listed for reference. The error is estimated using the covariance of
  $P_l$ at different $l$ and bins of $k$ for 1($h^{-1}$Gpc)$^3$
  boxes. The minimum chi-square values $\chi^2_{\rm min}$ in the
  fitting and the degree-of-freedom (d.o.f.) in parentheses are also
  listed. The unit of mass is $10^{13}h^{-1}M_\odot$.}
\label{tab:hod}
\end{center}
\end{table*}

\vspace{-0.5cm}
\subsection{Reconstructions of HOD from multipole power spectra}
We apply Markov Chain Monte Carlo method to the averaged simulated
power spectra over 100 realizations to estimate the likelihood
function of a parameter set $\mathbf{p}$ in the chi-square basis:
\begin{eqnarray}
\chi^2 &=&\sum_{i,j} [y_i^{\rm (model)}(\mathbf{p})-y_i^{\rm (sim)}]{\rm Cov}^{-1}_{ij}
[y_j^{\rm (model)}(\mathbf{p})-y_j^{\rm (sim)}] \nonumber \\
&&+[(N_{\rm tot}^{\rm (sim)}-N_{\rm tot}^{\rm (model)})/\sigma_N]^2,
\end{eqnarray}
where $y_i$ denotes $P_l(k)$ at each $l$ and bin of $k$, $\mathbf{p}$
include HOD parameters and the growth rate, and {\rm Cov} denotes the
covariance of $P_l(k)$ for 1($h^{-1}$Gpc)$^3$ boxes. The second term
on the right-hand side represents the constraint on the total number
of LRGs including both central and satellite LRGs. We give the error
of $N_{\rm tot}$ as $\sigma_N=\sqrt{N_{\rm tot}}$.

Table \ref{tab:hod} lists the best-fit values and the 1$\sigma$ errors
of the five HOD parameters in the functional form
(eq.[\ref{eq:HOD}]). The binning width $\Delta k$ is set to be
$0.016h$/Mpc and then the number of bins are 12 ($k\le 0.2h$/Mpc) and
37 ($k\le 0.6h$/Mpc) for each $P_l$. When the information of $P_0$ and
$P_2$ is included up to $k=0.2h$/Mpc, where the perturbation theory
agrees with the simulations \citep{Beutler13}, the addition of the
small-scale information of $P_4$ and $P_6$ reduces the error of HOD
parameters by nearly half.  When the information of $P_0$ and $P_2$ up
to $k=0.6h$/Mpc as same as $P_4$ and $P_6$ are included, the
improvement by adding $P_4$ and $P_6$ is 20-30\% for the satellite HOD
parameters of $M_{\rm cut}$ and $M_1$.  When varying the amplitude of
matter power spectrum ($\sigma_8$) and the residual shot noise $\delta
N$ as additional free parameters, the constraints only from $P_0$ and
$P_2$ becomes much weaker due to the degeneracy between the two-halo
term and the one-halo term. The information of $P_4$ and $P_6$, which
mainly determined by the one-halo term, helps to break the degeneracy
and significantly improve the HOD constraints.

\begin{figure}
\begin{center}
\includegraphics[width=8.4cm]{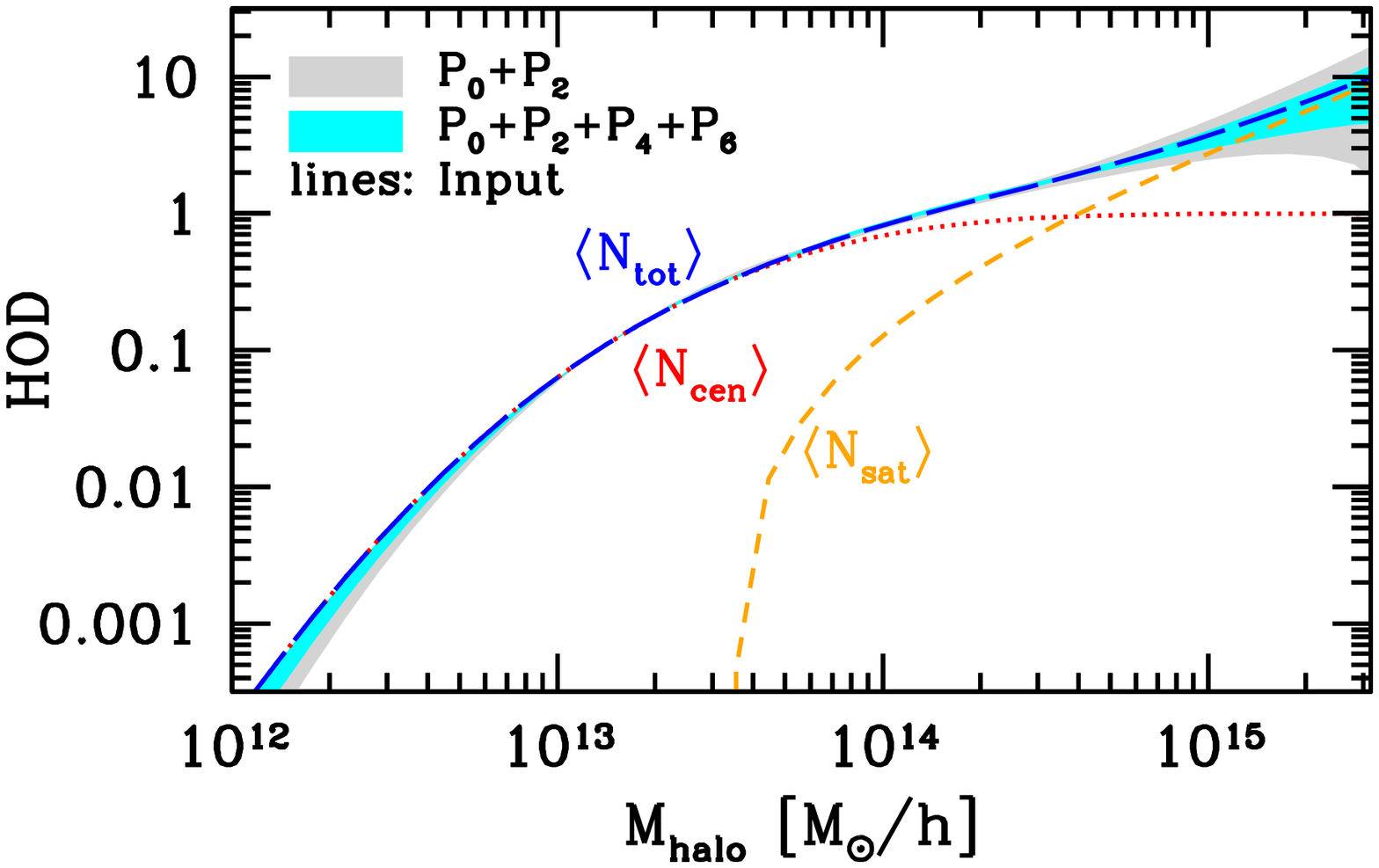}
\includegraphics[width=8.4cm]{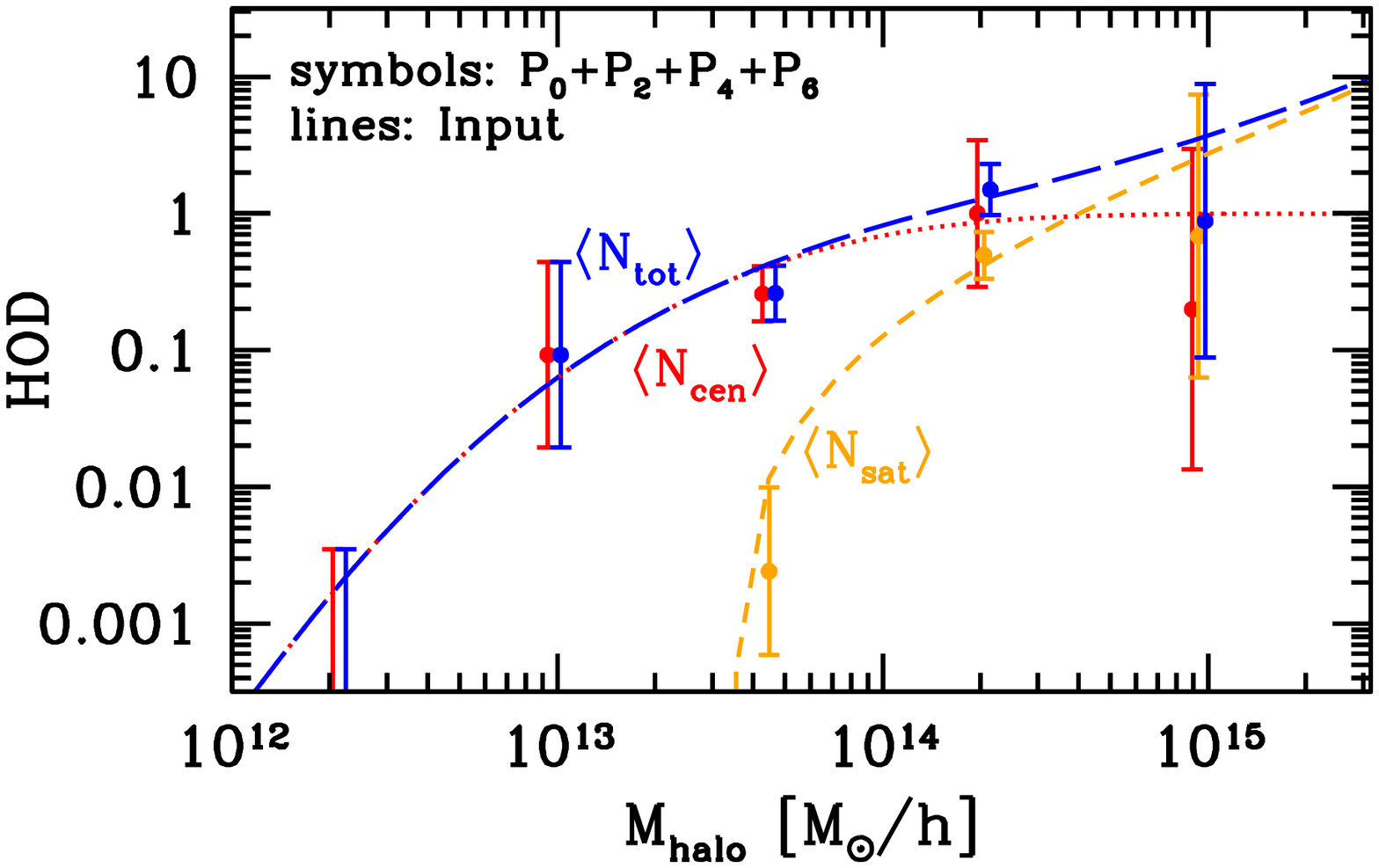}
\caption{Reconstructed HOD by fitting HOD parameters to $P_l(k)$ for
  the simulated LRG catalog. Upper panel shows the result when the
  functional form of HOD (eq. [\ref{eq:HOD}]) is assumed. Shaded area
  denote the 1-$\sigma$ error of $\langle N_{\rm tot}\rangle$ from
  $P_0$ and $P_2$ (Gray area) and $P_0, P_2, P_4$ and $P_6$ (light
  blue area). For reference, the input HOD for total (blue), central
  (red), and satellite galaxies (yellow) are plotted. Lower panel
  shows the result when central and satellite HOD values at 5
  different mass bins are directly measured without assuming any
  functional form using $P_0$, $P_2$, $P_4$ and $P_6$. The error-bars
  denote the 1-$\sigma$ error.}
\label{fig:hod}
\end{center}
\end{figure}

Figure \ref{fig:hod} shows the reconstructed HOD from $P_l$ of the
simulated LRG samples when the functional form of HOD is assumed
(upper panels) and when central and satellite HOD values is directly
fitted without assuming any functional form of HODs (lower
panels). The reconstructed HODs agree with the input HOD (lines)
within the 1-sigma error in both cases and the addition of high-$l$
multipole improve the HOD measurements as shown by the light-blue
shaded area in upper panel.

\vspace{-0.5cm}

\subsection{Constraints on the growth rate}
\label{subsec:4.3}
Fingers-of-God effect due to satellite galaxies is a major systematic
uncertainty in measuring the cosmic growth rate. High-$l$ multipoles
sensitive to the satellite properties improve the accuracy of growth
rate measurement \citep{HY2013}.  Table \ref{tab:fsat_gamma} shows the
constraints on the satellite fraction $f_{\rm sat}$ and the average
velocity dispersion $\sigma_v^{\rm (sat)}$ when using a functional
form of HOD and direct fitting of HOD values at 5 different bins of
mass. Note that $\sigma_v^{\rm (sat)}$ is determined by the HOD values
using the Virial theorem (eq.[\ref{eq:vsat}]). We find that addition
of $P_4$ and $P_6$ improve the measurement of the growth rate by
nearly twice (the error decreases from 8\% to 4\%) in both of the HOD
parametrization.

\begin{figure}
\begin{center}
\includegraphics[width=4.1cm]{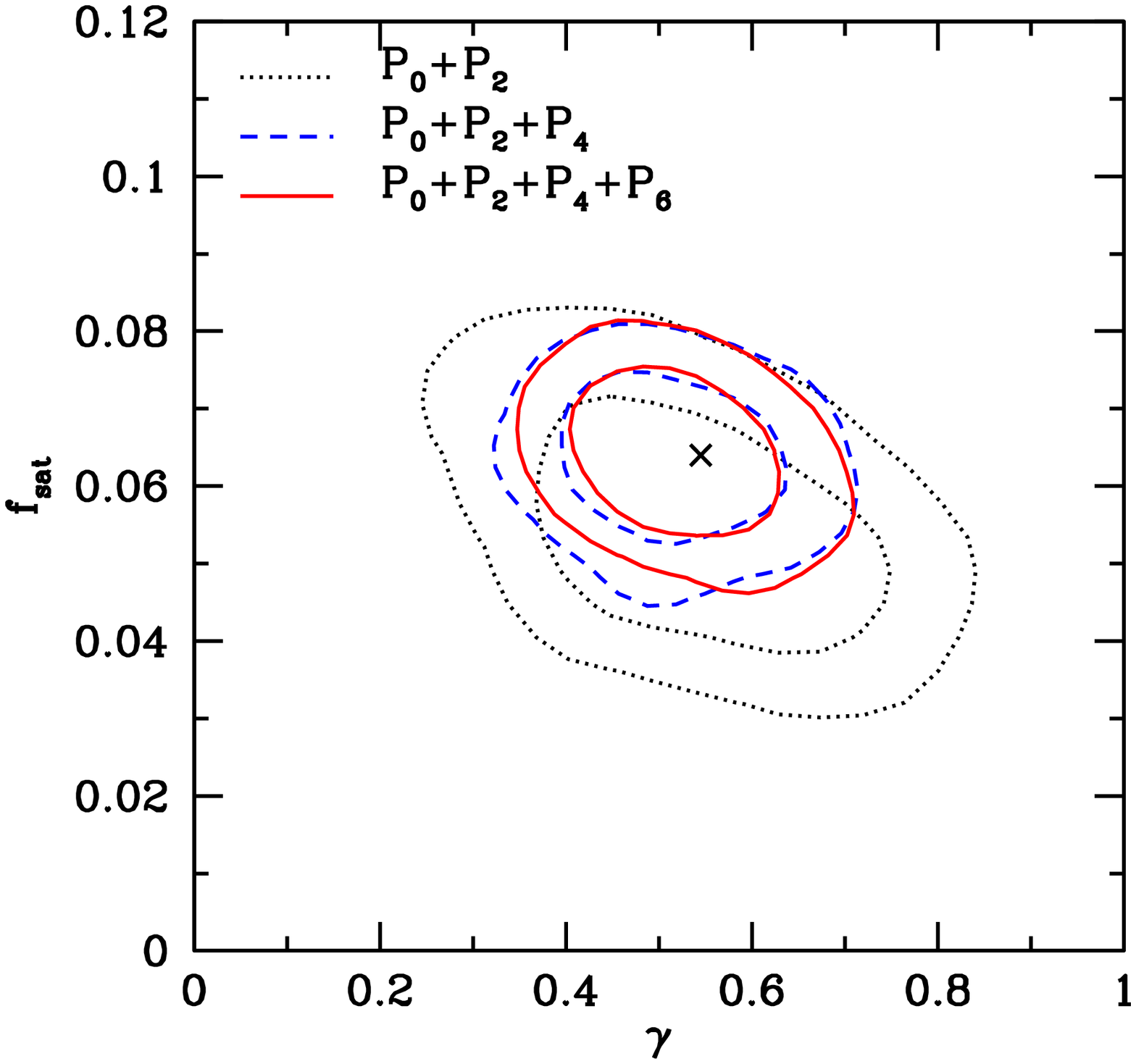}
\includegraphics[width=4.2cm]{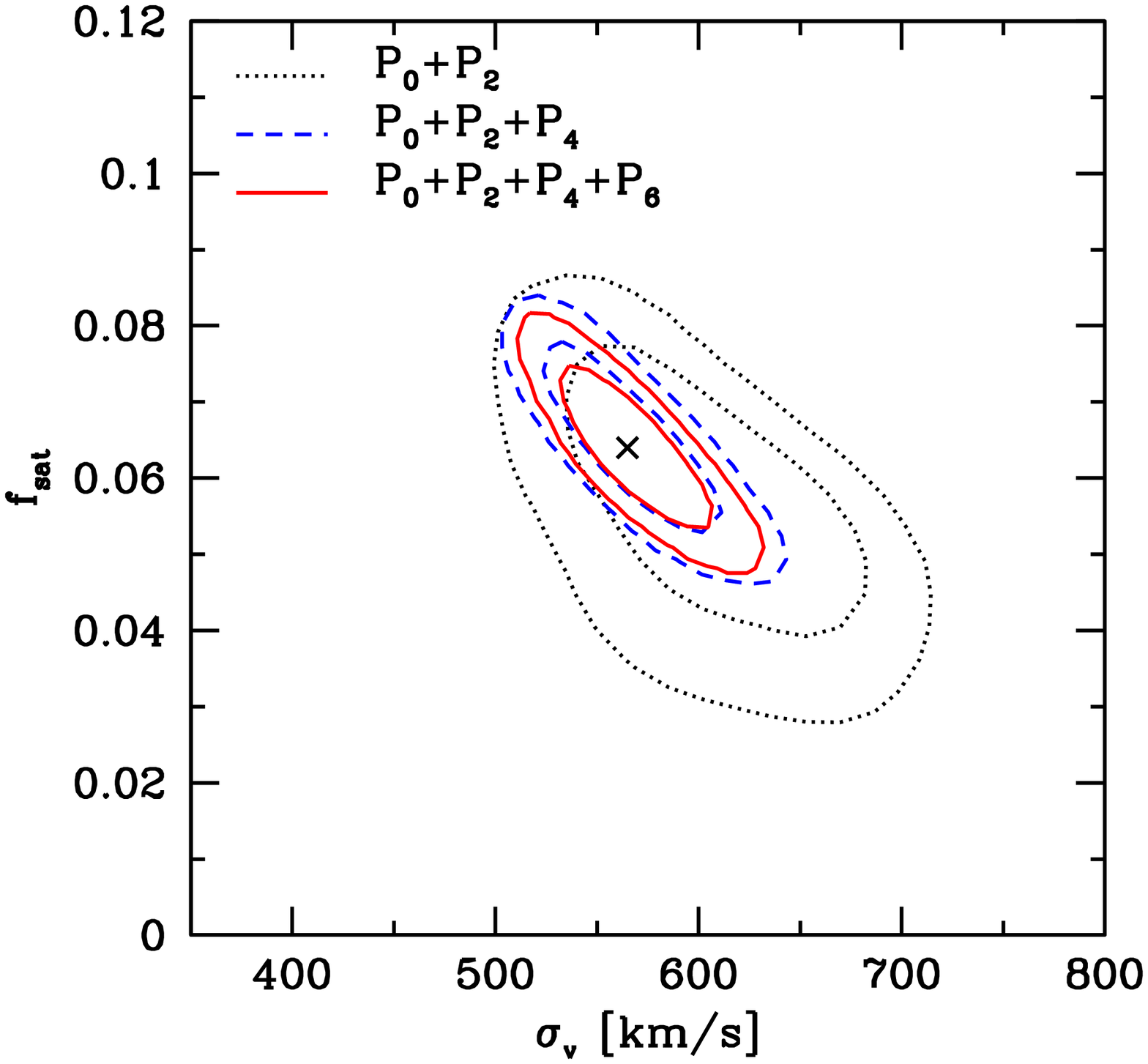}
\caption{Joint constraints on the satellite fraction $f_{\rm sat}$ and
  the growth rate index $\gamma$ (left) or $\sigma_v^{\rm (sat)}$
  (right) from $P_0+P_2$ (black), $P_0+P_2+P_4$ (blue), and
  $P_0+P_2+P_4+P_6$ (red). Here we use a functional form of HOD
  (eq [\ref{eq:HOD}]). Each contour denotes 68\% and 95\% error
  respectively. The symbol of crosses denote the input values.}
\label{fig:fsat_gamma}
\end{center}
\end{figure}

Figure \ref{fig:fsat_gamma} shows the joint constraints on the
fraction of satellite galaxies $f_{\rm sat}$ and the growth rate index
$\gamma$, which is calculated with the simple approximation
$f_z=\Omega_m^\gamma(z)$, and the satellite velocity dispersion
$\sigma_v^{\rm (sat)}$ from different combinations of $P_l$. The
parameters $f_{\rm sat}$ and $f_z$ degenerate with each other because
the suppression of quadrupole power due to satellite FoG effect is
mimicked by increasing the growth rate $f_z$ (or decreasing
$\gamma$). Figure \ref{fig:fsat_gamma} shows that the addition of
high-$l$ multipole measurements breaks their degeneracy and then
improves the accuracy of $\gamma$ by nearly twice. The input values
denoted by the cross symbols are successfully reproduced.

\begin{table}
\begin{center}
\begin{tabular}{ccccc}
\hline
\hline
parameters & input & $P_0,P_2$ & $P_0,P_2,P_4$ & $P_0,P_2,P_4,P_6$ \\
\hline
\multicolumn{5}{l}{HOD (5 parameter models in the functional form of eq. \ref{eq:HOD})} \\
\hline
$100f_{\rm sat}$   & 6.5   & $5.7\pm 1.2$   & $6.4 \pm 0.7$  & $6.4\pm 0.7$ \\
$\sigma_v^{\rm (sat)}$ [km/s]  & 570   & $604\pm 46$    & $570\pm 29$    & $569\pm 25$ \\
$f_z/f_z^{\rm GR}$ & 1     & $1.01\pm 0.09$ & $1.01\pm 0.06$ & $1.01\pm 0.05$ \\
$\gamma$           & 0.545  & $0.53\pm 0.12$ & $0.53\pm 0.08$ & $0.53\pm 0.07$ \\
\hline
\multicolumn{5}{l}{HOD (direct fitting of $\langle N_{\rm cen}\rangle$ and $\langle N_{\rm sat}\rangle$ at 5 mass bins)} \\
\hline
$100f_{\rm sat}$   & 6.5   & $5.9\pm 1.6$      & $6.0\pm 0.8$   & $6.2\pm 0.7$ \\
$\sigma_v^{\rm (sat)}$ [km/s]  & 570   & $583\pm 86$    & $556\pm 40$    & $571\pm 26$  \\
$f_z/f_z^{\rm GR}$ & 1     & $1.03\pm 0.09$ & $1.03\pm 0.06$ & $1.01\pm 0.05$ \\
$\gamma$           & 0.545 & $0.51\pm 0.12$  & $0.51\pm 0.08$ & $0.53\pm 0.07$ \\
\hline
\end{tabular}
\caption{Constraints on the satellite fraction $f_{\rm sat}$, the
  satellite velocity dispersion $\sigma_v^{\rm (sat)}$, the growth
  rate normalized by the GR predictions $f_z/f_z^{\rm GR}$ and the
  growth rate index $\gamma$ from different combinations of $P_l$. We
  use $k\le 0.2$ for $P_0$ and $P_2$ and $k\le 0.6$ for $P_4$ and
  $P_6$. We show the case when using a functional form of HOD in the
  equation \ref{eq:HOD} (Upper) and when fitting central and satellite
  HOD values at 5 different bins of halo mass as free parameters but
  assuming that $\langle N_{\rm sat}\rangle$ increases as larger $M$
  (Lower).
\label{tab:fsat_gamma}
}
\end{center}
\end{table}

\vspace{-0.5cm}

\section{Summary and Conclusions}
\label{sec:summary}

We present the HOD constraints from multipole galaxy power spectra
$P_l(k)$ using the simulated catalogs which follows HOD of SDSS LRG
samples.  The high-$l$ multipole spectra such as $P_4$ and $P_6$ are
sensitive to the Fingers-of-God (FoG) effect due to the large internal
motion of satellite galaxies and thus they are useful probe to
constrain the fraction and the velocity dispersion of satellite
galaxies.  We find that the input HOD is successfully reconstructed
from $P_l$ and that high-$l$ multipole spectra significantly improve
the accuracy of the HOD measurements. We also find that the addition
of high-$l$ multipole information improve the accuracy of the growth
rate measurement by nearly twice because they break the degeneracy
between the cosmic growth rate and the satellite FoG.

In this letter we simply use the simulated samples in $1(h^{-1}{\rm
  Gpc})^3$ cubic box. To apply our method for the actual observations,
we need to consider the various observational issues such as the
angular mask, radial selection function and the fiber collision. Such
observational effect can be evaluated by making the mock samples with
the same geometry as the observation. Various methods to correct the
fiber collision effect has been also developed
\citep[e.g.,][]{Guo12}. We also need to consider the uncertainty of
cosmology and halo power spectra more exactly. The uncertainty mainly
affect $P_0$ and $P_2$ where the two-halo term is dominant, while
high-$l$ multipole power spectra dominated by the one-halo term are
insensitive to the details of the halo power spectrum. Even when
fitting cosmology parameters together with HOD parameters, high-$l$
multipole spectra should be still important to eliminate the satellite
FoG effect.  Measuring the deviation of the satellite velocity
dispersion from the virial velocity dispersion provides useful
information related to the satellite kinematics
\citep[e.g.,][]{Tinker07}. These works are beyond the scope of this
letter and left for the future.

\vspace{-0.5cm}

\section*{Acknowledgments}
We thank anonymous reviewer for careful reading and useful comments.  
We also thank K. Yamamoto for useful discussions.  The research is
supported by Grant-in-Aid for Scientific researcher of Japanese
Ministry of Education, Culture, Sports, Science and Technology
(No.~24740160).

\vspace{-0.5cm}

\bibliographystyle{mn2e} 

\label{lastpage}

\end{document}